# Magnetotransport and the upper critical magnetic field in MgB$_2$

P. Szabó[a,c], P. Samuely[a], A.G.M. Jansen[b], T. Klein[c], J. Marcus[c], D. Fruchart[d],

S. Miraglia[d]

[a]*Institute of Experimental Physics, SAS Kosice, Slovakia*

[b]*Grenoble High Magnetic Field Laboratory, MPI-FKF and CNRS, France*

[c]*LEPES CNRS, Grenoble, France*

[d]*Laboratoire de Cristallographie, CNRS, Grenoble, France*

**Abstract**

Magnetotransport measurements are presented on polycrystalline MgB$_2$ samples. The 'resistive' upper critical magnetic field reveals a temperature dependence with a positive curvature from $T_c = 39.3$ K down to about 20 K, then changes to a slightly negative curvature reaching 25 T at 1.5 K. The 25-Tesla upper critical field is much higher than what is known so far on polycrystals of MgB$_2$ but it is in agreement with recent data obtained on epitaxial MgB$_2$ films. The deviation of $B_{c2}(T)$ from standard BCS might be due to the proposed two-gap superconductivity in this compound. The observed quadratic normal-state magnetoresistance with validity of Kohler's rule can be ascribed to classical trajectory effects in the low-field limit.

*Contact address: pszabo@saske.sk*

## 1. Introduction

Since the discovery of superconductivity in magnesium diboride at 39 K in January of this year the study of the physical properties in this intermetallic compound has aroused an enormous interest. A lot of preprints are appearing on the Los Alamos cond-mat server concerning the transport properties of MgB$_2$. The results are still sample dependent and controversial. The reported values of the residual resistivity differ by a factor of 1000 starting from 0.3 µΩ cm [1-3] and the residual resistivity ratio RRR ranges from 1 up to 25 [1,2,4,5]. Different magnitudes of the normal state magnetoresistance have been observed [2-5]. The zero-temperature value of the upper critical magnetic field $B_{c2}(0)$ has been found mostly around 15 T [6]. Recently, in epitaxial films a value of about 30 Tesla for $B_{c2}(0)$ has been reported with a small anisotropy [4]. The temperature dependence of $B_{c2}$ reveals usually some deviations from the conventional Werthamer-Helfand-Hohenberg dependence [7] showing a non-quadratic temperature dependence at low temperatures and a positive curvature at higher temperatures up to $T_c$. Shulga *et al.* [8] ascribed these deviations to the proposed theoretical model of two-band superconductivity in MgB$_2$ [9]. In our previous experiments on Andreev-reflexion spectroscopy in MgB$_2$ [10], we have shown in a direct way the presence of two superconducting energy gaps up to the same bulk transition temperature $T_c$ indicating that the two gaps are inherent to the superconductivity in MgB$_2$. We present here a magnetotransport study on MgB$_2$ polycrystals from the same badge as used in our previous spectroscopic experiments in order to get information on the temperature dependence of the upper critical magnetic field and on the normal-state magnetoresistance.

## 2. Experiment

Transport experiments have been performed on polycrystalline MgB$_2$ samples. Four electrical contacts for the resistivity measurements have been prepared by silver painting the freshly cleaned corners of the top side of the sample in a Van der Pauw's configuration. A standard phase-sensitive detection technique at 17 Hz was used to measure the temperature and magnetic field dependence of the resistance. All measurements were performed with the magnetic field applied perpendicularly to the excitation current. The field was generated by a 28 Tesla resistive magnet in the Grenoble High Magnetic Field Laboratory.

The MgB$_2$ samples were prepared from boron powder (99,5 % pure, Ventron) and magnesium powder (98% Mg + 2 % KCl, MCP Techn.), in relative proportion 1.05 : 2. A 2 g mass of the mixed powders was introduced into a tantalum tube, then sealed by arc melting under argon atmosphere (purity 5N5). The tantalum ampoule was heated by high frequency induction at 950°C for about 3 hours. After cooling down to room temperature, the sample was analysed by X-ray diffraction and Scanning Electron Microscope. Among the brittle dark grey MgB$_2$ powder (grain size < 20 μm), a few larger grains (0.1 to 1 mm) were found. Laue patterns show evidence for only a limited number of single crystals in each grain. Resistivity and a.c. susceptibility measurements of these larger grains reveal a particularly abrupt transition indicating their high quality in comparison with that of the fine powder. The dimension of the investigated platelet-shaped grain is about 0.3×0.4×0.03 mm$^3$. The applied current density is about 3 A/cm$^2$. The superconducting resistive transition is very narrow with a midpoint $T_c$ = 39.3 K and a width $\Delta T_c$ = 0.6 K, defined as the difference in the temperatures between the 90% and 10% values of the normal state resistivity. In the normal state the resistivity $\rho$ increases from 5 μΩcm up to the room temperature value 20 μΩcm, yielding a residual resistivity ratio RRR = $\rho$(300 K) / $\rho$(40 K) = 4.

## 3. Results and discussion

Figure 1 displays our magnetotransport data measured at different fixed temperatures from $T$ = 1.5 K up to $T_c$. The full set of the resistivity data in a magnetic field shows conventional superconducting transitions, which are shifted to higher fields and gradually broadened with decreasing temperatures from $\Delta B$=0.3 T at $T$ = 36 K to $\Delta B$ = 10 T at $T$ = 1.5 K. The resistivity $\rho(H)$ measured in the normal state at 50 K follows exactly the curve taken at 39 K apart from the drop to the superconducting state at very low magnetic fields. The normal state magnetoresistivity $\Delta\rho(H)/\rho(0) = [\rho(H) - \rho(0)]/\rho(H)$ equals 25% at 28 T.

In Figure 2 the normal state magnetoresistivity $\Delta\rho(H)/\rho(0)$ measured at $T$ = 50 K is shown in two different double-logarithmic scales. On the double-logarithmic scale, the field dependence of $\Delta\rho(H)/\rho(0)$ is a straight line with a slope equal to 2 up to a magnetic field of about 12 Tesla. At higher fields a deviation (emphasized in the inset) from this quadratic magnetoresistance can be observed. The quadratic magnetoresistance would be expected for classical trajectory effects in the low-field limit $l/r_c < 1$ (mean free path $l$ and cyclotron radius $r_c = mv_F/eB$ with electron mass $m$ and Fermi velocity $v_F$) where $\Delta\rho(H)/\rho(0) \gg a(l/r_c)^2$ with the prefactor $a$ depending on the details of the Fermi surface.

Chen et al. [11] have found a correlation between the residual resistivity ratio RRR and the value of the normal-state magnetoresistivity $\Delta\rho(5T)/\rho(0) = 0.04 \, (RRR)^{2.2}$ at 50 K and under 5 T for four different samples with RRR ranging from 1.16 up to 8.3. Our data with $\Delta\rho(H)/\rho(0) = 0.9\%$ at $B$= 5 T and RRR = 4 fit to this formula very well. In fact, this formula with a nearly quadratic dependence on RRR, i.e. on the inverse of the low temperature resistivity $\rho(0)$, can be understood if only classical trajectory effects are affecting the magnetoresistance such that the Kohler relation $\Delta\rho(H)/\rho(0) = \text{const.}[B/\rho(0)]^2$ holds. Above 12 Tesla we observe a small deviation from the quadratic magnetoresistance towards a more linear field dependence at higher fields.

In Fig. 3 the 'resistive' critical magnetic field values determined from our magnetotransport data are plotted. The temperature dependence of the resistive critical magnetic fields $B^*(T)$ has been determined for different criteria, i.e. $B^*_{onset}$ at the onset of resistivity (5% of the normal state resistivity), $B^*_{mid}$ at 50% of the normal state resistivity, and $B_{c2}$ at the end of the transition (crossing with the normal state resistivity). The large width of the resistive transitions ($B_{c2}(T) - B^*_{onset}(T)$) at all temperatures suggests the importance of dissipation processes due to depinned vortices. It is worth noting that $B^*_{onset}(T)$ line is almost the same in all reported measurements [6]. Obviously, for practical applications this onset field is the most important one. It also coincides well with the irreversibility field $B_{irr}(T)$. Recently, by alloying with oxygen the MgB$_2$ thin films with artificially introduced pinning centers achieved an irreversibility field bigger than 14 T at 4.2 K [12]. The value of the mean field transition $B_{c2}(0) = 25$ T represents to our knowledge the highest value of the upper critical magnetic field measured on MgB$_2$ bulk samples [6]. Similarly high values of $B_{c2}$ have been published only on epitaxial MgB$_2$ films [4]. The coherence length can be calculated from the Ginzburg-Landau relation $\xi = [\Phi_0/2\pi B_{c2}(0)]^{1/2}$. Substituing $B_{c2}(0)=25$ T and flux quantum $\Phi_0 = h/2e$ we find $\xi = 3.6$ nm. A value of the mean free path $l$ can be calculated from the Drude formula for the normal state resistivity $l = mv_F / ne^2 \rho(40\,K)$ using a free electron mass $m$ for the quasiparticles, Fermi velocity $v_F \sim 4.8 \times 10^7$ cms$^{-1}$ and charge carrier density $n \sim 6.7 \times 10^{22}$ cm$^{-3}$ [1]. The resulting mean free path value is $l \sim 5.0$ nm. From the above estimated value of the superconducting coherence length follows that our polycrystalline MgB$_2$ sample satisfies the clean limit condition $l > \xi$. This opens a way for possible applications, because when additional pinning centers will be introduced into the pure sample, the vortex pinning will shift the onset of the resistance or irreversibility line up to higher magnetic fields.

It is important to notice, that all three critical fields $B^*(T)$ obtained from our magnetotransport data give the same type of temperature dependence with a negative curvature at low temperatures and then changing to a positive curvature above 20 K up to $T_c$. For high-$T_c$ superconductors it is often observed that the positive curvature changes to an overall negative curvature by taking resistivity criteria for the transition field more close to the normal state. The classical WHH theory predicts a linear dependence near $T_c$ with saturation at the lowest temperatures. Shulga et al. [8] proposed to describe this unconventional behavior by considering the model of two-band/two-gap superconductivity. We have found two distinct superconducting gaps with the same $T_c$ on the same sample in a direct spectroscopic Andreev reflection experiment [10].

## 4. Conclusion

Magnetotransport experiment have been performed on the polycrystalline MgB$_2$ sample. The upper critical magnetic field has been determined from our data. Its 25-Tesla value at zero temperature represents to the best of our knowledge the highest value measured on bulk MgB$_2$ samples. The estimated values of the superconducting coherence length and mean free path indicate, that our sample is in the clean limit, what is stimulating for possible applications at high magnetic fields. The temperature dependence of the upper critical magnetic field reveals upward curvature near $T_c$.

**Acknowledgments**

Support of the Slovak VEGA grant No.1148, of the HPP Programme "Transnational Access to Major Research Infrastructures" (P.S., P.Sz.), and of the Deutscher Akademischer Austauschdienst (P.S.) is greatly acknowledged.

**Figure captions:**

Figure 1.: Magnetoresistivity measured up to magnetic fields $B$ = 28 T at the indicated temperatures.

Figure 2.: The normal state magnetoresistivity (open circles) measured at $T$ = 50 K shown in two different double-logarithmic scales. The straight line has a slope of 2 corresponding to a quadratic magnetoresistance.

Figure 3.: Temperature dependencies of the 'resistive' critical magnetic fields, determined from the resistive transitions shown in Figure 2 using different criteria for the transition fields. Open circles show $B^*_{onset}(T)$ at the onset of the superconducting transition (determined at 5% of the normal state resistivity), solid circles $B^*_{mid}(T)$ for the mid-point field (50% of the normal state resistivity) and open squares $B_{c2}(T)$ for the upper critical field (crossing point with the normal state resistivity).

Figure 1:

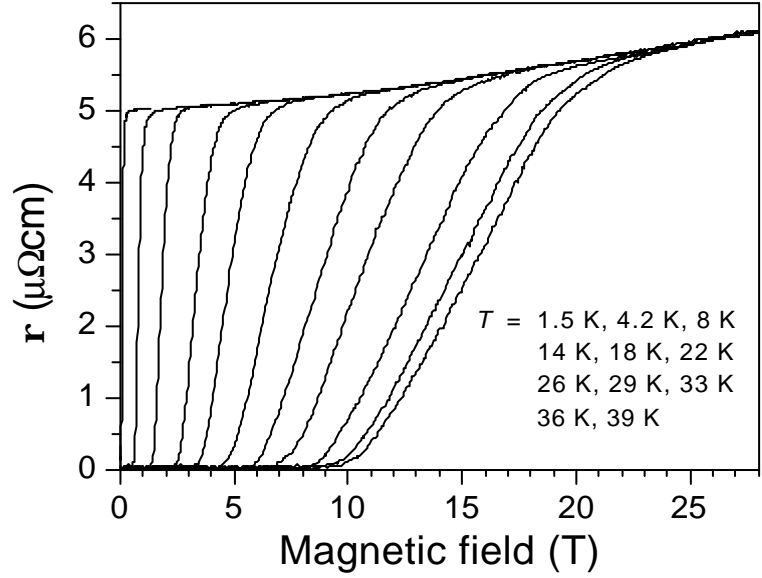

Figure 2:

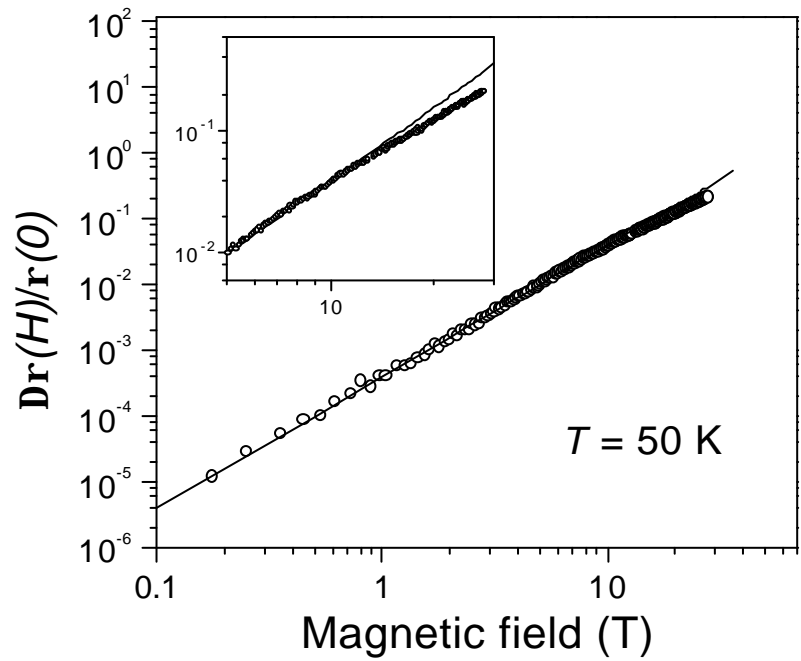

Figure 3:

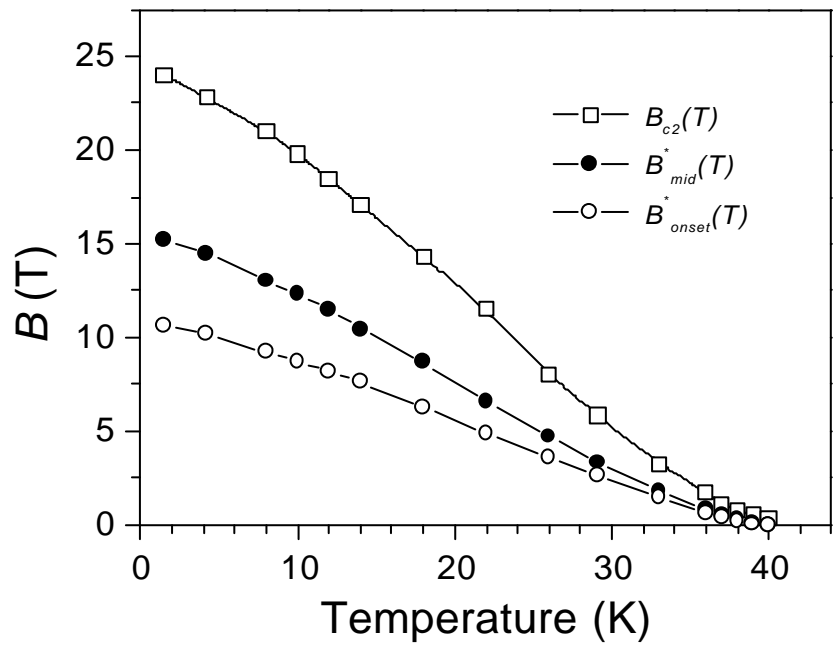